\documentclass[12pt,a4paper]{article}

\usepackage{CJK}
\usepackage{amsmath}
\usepackage{amsfonts}
\usepackage{amssymb}
\usepackage{graphicx}
\usepackage{color}

\newcommand{\be}{\begin{eqnarray}}
\newcommand{\ee}{\end{eqnarray}}

\newcommand{\BE}{\mathbf{E}}

\newcommand{\BF}{\mathbf{F}}

\newcommand{\BB}{\mathbf{B}}

\renewcommand{\bar}{\overline}

\newcommand{\p}{\partial}
\newcommand{\half}{\frac{1}{2}}

\begin{document}

\begin{center}
{\Large \textbf{Solving Maxwell's Equations}}\\
\bigskip
\smallskip
\centerline{Tong Chern$^\dagger$}
\smallskip
\centerline{\it School of Science, East China University of Technology, Nanchang 330013, China}
\medskip \centerline{\it $^\dagger$ tongchen@ecut.edu.cn}
\end{center}

\bigskip

\begin{abstract}
This paper discusses the use of the Riemann-Silberstein vector to solve the source-free Maxwell's equations
and obtains novel analytical solutions.
The solving process naturally leads to the spinor form
of the source-free Maxwell's equations.
Several powerful theorems are established to solve this spinor form equation.
The Waveguide Solution Theorem provides an elegant way to solve waveguide problems,
while The Schrodinger Solution Theorem connects the Maxwell's equations
with the two-dimensional Schrodinger equation.
By utilizing The Schrodinger Solution Theorem,
a precise formula for spatiotemporal diffraction of the Maxwell's equations is derived,
which allows for the reconstruction of electromagnetic waves throughout space and time
based on the field distribution on the diffraction screen.
And finally, by studying some relevant contour integrals, two additional simple but beautiful mathematical theorems that are necessarily satisfied by electromagnetic field in any source-free region are derived.
\end{abstract}
\bigskip
[\textbf{In this paper, we take the unit system of $\epsilon_0=\mu_0=c=1$.}]\\

\section{Introduction}

By introducing the complex vector field $\mathbf{F}=\mathbf{E}+i\mathbf{B}$(known as the Riemann-Silberstein vector),
where $\mathbf{E}$ stands for the electric field intensity and $\mathbf{B}$ stands for the magnetic field intensity,
and $i$ is the imaginary unit, the source-free Maxwell's equations can be rewritten in the following form,
\begin{align} & i\p_t\BF =\nabla\times\BF,\label{m1}\\
&\nabla\cdot\BF =0.\label{m2}\end{align}
This form of the Maxwell's equations is useful both at the classical and quantum levels
\cite{RS}. However, its analytical solution is not straightforward to obtain,
and some indirect methods have been proposed in the literature\cite{RS,Birula}.
In contrast, the starting point of this paper is to directly solve equations (\ref{m1}) and (\ref{m2})
under certain simplifying assumptions. In deed, this paper successfully
obtains several novel analytical solutions of the source-free Maxwell's equations.

Although the concept of spinor
is not necessary for our purposes, the solving process naturally leads us to the spinor form of the source-free Maxwell's equations, as previously derived by Penrose \cite{Penrose1}. This form was traditionally solved through the highly abstract twistor transformation
\cite{Penrose2,Twistor}. While the twistor transformation is elegant in its form,
in practical applications,
people are more interested in the specific form
of the solutions rather than their integral representations, and
the twistor transformation is inconvenient to use and too abstract to be useful.

In contrast, this paper establishes a number of powerful
ground-based theorems for solving this spinor form of Maxwell's equations.
The General Theorem relates the solutions of Maxwell's equations to the solutions of the scalar D'Alembert equation,
and it tells us that any solution of the scalar D'Alembert equation gives a class of solutions of Maxwell's equations
by simple univariate indefinite integration. Using this General Theorem,
one can in turn derive what we call The Waveguide Solution Theorem,
which tells us that any solution of the two-dimensional Helmholtz equation gives a class of solutions of Maxwell's equations, and this theorem provides an elegant way of solving the waveguide problem.
Even more interesting is what we call Schrodinger Solution Theorem,
which establishes a connection between Maxwell's equations and the two-dimensional Schrodinger equation;
specifically, it tells us that any solution of the two-dimensional free-particle Schrodinger equation
gives a class of solutions of the source free Maxwell's equations by simple univariate indefinite integration.
Using Schrodinger Solution Theorem, we derive a Spacetime Diffraction Formula for Maxwell's equations,
which is an exact formula for reconstructing electromagnetic waves throughout spacetime using field distributions on a given diffraction screen!

And in the last section of this paper, 
by studying some contour integrals of the electromagnetic fields, two additional simple but beautiful mathematical theorems that are necessarily satisfied by electromagnetic field in any source-free region are derived.

\section{Solutions for the simple case}

In order to specifically
solve equation (\ref{m1}) and equation
(\ref{m2}). We write down the component form of $\BF$ as $\BF=(F_x, F_y, F_z)$.
And we need to start with some simplifying assumptions.

First, examine the simplest case by assuming \be F_z=0.\label{solution1}\ee
Let's find the solution to the system of equations (\ref{m1}),(\ref{m2}) that satisfies this assumption.
It is not difficult to obtain that the component
form of the equation (\ref{m1}) takes \begin{align} & i\p_tF_x=-\p_zF_y,\label{mf1}\\
& i\p_tF_y=\p_zF_x,\label{mf2}\\
& \p_xF_y-\p_yF_x=0.\label{mf3}\end{align}
And the equation (\ref{m2}) becomes
\be \p_xF_x+\p_yF_y=0.\label{mf4}\ee

Next, introduce the complex coordinates $w=x+iy,\bar{w}=x-iy$ in the $(x,y)$ plane. Thus we have $\p_w=\half(\p_x-i\p_y)$, $\p_{\bar{w}}=\half(\p_x+i\p_y)$.
Further define \be F_x+iF_y=F_{\bar{w}},\quad F_x-iF_y=F_{w},\ee
note that since $F_x, F_y$ are complex fields,
thus $F_w$ and $F_{\bar{w}}$ are two complex fields independent of each other,
and are not complex conjugates of each other.
And since $F_z=0$ is assumed, it is not hard to see that there is now $\BF^2=F_wF_{\bar{w}}$.

Using the above definitions, equation (\ref{mf1}) and equation (\ref{mf2}) can be recombined as
\begin{align} (\p_t + \p_z)F_{w}=0,\quad (\p_t-\p_z)F_{\bar{w}}=0. \label{mcf1} \end{align}
and the equations (\ref{mf3}) and (\ref{mf4}) can be rewritten as
\be \p_{w}F_{\bar{w}}-\p_{\bar{w}}F_w=0,\quad \p_wF_{\bar{w}}+\p_{\bar{w}}F_w=0.\ee
Adding and subtracting these two equations gives us
\be \p_{w}F_{\bar{w}}=0,\quad \p_{\bar{w}}F_w=0.\label{mcf2}\ee

Equation (\ref{mcf1}) tells us that $F_w=f(t-z, w,\bar{w})$, $F_{\bar{w}}=g(t+z, w, \bar{w})$, where $f, g$ are two arbitrary complex-valued functions. And the equation (\ref{mcf2}) tells us that $F_w$ is a holomorphic function of $w$ and $F_{\bar{w}}$ is an anti-holomorphic function of $\bar{w}$, thus we have \be F_w=f(t-z, w)\quad F_{\bar{w}}=g(t+z,\bar{w}). \label{solution2}\ee

The $F_z=0$ plus the above results (\ref{solution2}) consists
a large class of analytical solutions to the vacuum
source-free Maxwell's equations,
and it is clear that in general it describes the superposition of
left- and right-moving waves along the $z$-axis (treating the $z$-axis as a horizontal axis),
and since $F_z=0$, it can be seen that the polarizations of these waves are perpendicular to the propagation direction, thus there are transverse waves!
And, it might be useful to point out here that the electrodynamics textbook approach\cite{Feynman} on how to solve electrostatic and magnetostatic problems in the two-dimensional $(x,y)$ plane by using holomorphic functions
can be viewed as a special case of the above solution.

Since the above solution describes
a linear superposition of a left traveling wave and a right traveling wave,
it may be useful to consider only the right traveling wave,
it is equivalent to taking the solution as
\be F_z=0,\quad F_{\bar{w}}=0, \quad F_w=f(t-z, w). \label{solution3}\ee
This solution clearly has $\BF^2=F_wF_{\bar{w}}=0$, and since $\BF^2=\BE^2-\BB^2+i2\BE\cdot\BB$,
it implies that $|\BE|=|\BB|$, and $\BE$ and $\BB$ are mutually orthogonal.
So for this right traveling wave solution,
$|\BE|=|\BB|$, and the electric and magnetic fields are mutually orthogonal!

We can write down the $x,y,z$ components of the above right traveling wave solution \begin{align} F_x &=\half(F_{w}+F_{\bar{w}})=\half f(t-z, w)\nonumber\\
F_y &=i\half (F_{w}-F_{\bar{w}})=i\half f(t-z, w)\nonumber\\
F_z &=0.\nonumber \end{align}
It is then easy to separate the real and imaginary parts
and find the electric and magnetic field components respectively.

In fact, the above right traveling wave solution
was first discovered in the literature \cite{vortex}
by a different method.
This solution contains a rich vortex structure,
and the locations of these vortices in the $(x, y)$ plane can be determined
from the zeros of the holomorphic function $f(t-z, w)$, i.e., from the factorization of $f(t-z, w)$ into
\be f(t-z, w)=(w-a_1(t-z))(w-a_2(t-z))... (w-a_n(t-z)). \ee
Where $a_i(t-z)$ are complex functions of the variable $t-z$.

Another possible application of the above solution is that,
assuming that there are sources in the region near the origin of the $(x, y)$ plane,
and that the region away from the origin is source-free,
then our right traveling-wave solution (\ref{solution3})
is only applicable to the region away from the origin of the $(x, y)$ plane (and an appropriate solution of
the Maxwell's equations with sources is to be spliced in near the origin.
How to splice in both solutions is outside the scope of this paper.
For the two-dimensional electrostatic and magnetostatic cases,
this problem has been examined in the literature\cite{source}).
Therefore, at this point, $f(t-z, w)$ can be Laurent expanded at the origin of $w=0$,
thus we have \begin{align} f(t-z, w)=&\frac{c_{-m}(t-z)}{w^m}+\frac{c_{-m+1}(t-z)}{w^{m-1}}+... +\frac{c_{-1}(t-z)}{w}\nonumber\\
&+c_0(t-z)+c_1(t-z)w^1+c_2(t-z)w^2+.... \end{align} where $c_n(t-z)$ are arbitrary complex functions of $t-z$. Physically, it is often required that the field configuration tends to zero at $|w|\rightarrow +\infty$, thus in order to satisfy this requirement one has to take \begin{align} f(t-z, w)=\frac{c_{-m}(t-z)}{w^m}+\frac{c_{-m+1}(t-z)}{w^{m-1}} +... +\frac{c_{-1}(t-z)}{w}.
\end{align}

\section{More analytical solutions}

In the following, we will move to the case of $F_z\neq 0$.
First, inspired by the above solution,
we further introduce the light-cone coordinates $v=t+z$, $u=t-z$
(thus $\p_v=\half (\p_t+\p_z)$, $\p_u=\half (\p_t-\p_z)$).
It is not difficult to verify that the complete vacuum source-free Maxwell's
equations can be written in the following equivalent form:
the $x, y$ components of equation (\ref{m1}) can be recombined as
\begin{align}&\p_v F_w=\p_wF_z,\label{m1c1}\\
&\p_u F_{\bar{w}}=-\p_{\bar{w}} F_z.\label{m1c2}\end{align}
And the $z$ component of equation (\ref{m1}) can be rewritten as
\be (\p_{\bar{w}}F_w-\p_wF_{\bar{w}})-\p_tF_z=0.\label{m1c03}\ee
On the other hand, equation (\ref{m2}) can be written as
\be (\p_{\bar{w}}F_w+\p_wF_{\bar{w}})+\p_zF_z=0.\label{m1c04}\ee
Adding and subtracting the equations (\ref{m1c03}) and (\ref{m1c04})
will give us \begin{align}&\p_{\bar{ w}}F_w=\p_uF_z,\label{m1c3}\\
&\p_wF_{\bar{w}}=-\p_vF_z.\label{m1c4}\end{align}
The equations (\ref{m1c1}), (\ref{m1c2}), (\ref{m1c3}), (\ref{m1c4})
are the equivalent forms of the vacuum source-free Maxwell's equations.

In order to get solutions with nice form
of the above equations,
we still need some assumptions,
however, we now slightly relax the previous assumption
of $F_z=0$.
This time we assume \be F_z=C(v, w). \ee
Here the function $C$ is a holomorphic function with respect to $w$.
That is, we assume that $\p_uF_z=0$, $\p_{\bar{w}}F_z=0$.
Obviously, under this assumption,
equations (\ref{m1c1}) and (\ref{m1c2}) become \begin{align} &\p_vF_w=C_w(v, w),\label{m1c11}\\
&\p_u F_{\bar{w}}=0\label{m1c22}\end{align} respectively,
where $C_w$ denotes the first-order partial derivative of the function $C$ with
respect to the independent variable $w$.
Integrating this set of equations gives its
general solution $F_w=\int dv C_w(v, w)+f(u, w, \bar{w})$, $F_{\bar{w}}=g(v, w,\bar{w})$.
And the equation (\ref{m1c3}) now becomes \be\p_{\bar{w}}F_w=0.\ee
So we have \be F_w=\int dv C_w(v, w)+f(u, w). \ee where
$f$ is a holomorphic function with respect to $w$.
However, the equation (\ref{m1c4}) now becomes
\be \p_wF_{\bar{w}}=-C_v(v, w),\ee where $C_v$ represents
the first-order partial derivative of the function $C$ with respect to the independent variable $v$.
By integrating this equation, it is easy to see that its general
solution is \be F_{\bar{w}}=-\int dw C_v(v, w)+g(v,\bar{w}).
\ee Where $g(v,\bar{w})$ is an anti-holomorphic function of $\bar{w}$.

In summary, we obtain the following analytical
solution to the source-free Maxwell's equations \begin{align}F_z&=C(v, w),\nonumber\\
 F_w&=\int dv C_w(v, w)+f(u, w),\nonumber\\
F_{\bar{w}}&=-\int dw C_v(v, w)+g(v,\bar{w}). \nonumber\end{align}
Here, the integrals are all indefinite integrals.

In exactly the same way, one can discuss the situation of $F_z=T(u,\bar{w})$
($T$ is an arbitrary function of the independent variables, provided it is anti-holomorphic with respect to $\bar{w}$).
Of course, we can also use the principle of superposition to superimpose the solutions of these two cases ($F_z=C(v, w)$ and $F_z=T(u,\bar{w})$).
The final solution can be obtained as follows,\begin{align}F_z&=C(v, w)+T(u,\bar{w}),\nonumber\\
 F_w&=\int dv C_w(v, w)+\int d\bar{w}T_u(u,\bar{w})+f(u, w),\nonumber\\\
F_{\bar{w}}&=-\int dw C_v(v, w)-\int du T_{\bar{w}}(u,\bar{w})+g(v,\bar{w}). \label{analytics1}\end{align}
where $T_u\equiv\p_u T$, $T_{\bar{w}}\equiv\p_{\bar{w}}T$.

This final solution obviously depicts an electromagnetic
wave propagating along the $z$-axis, however,
the direction of the electromagnetic field is not perpendicular to
the direction of propagation because of $F_z\neq 0$!
Although in the plane wave solution of Maxwell's equations,
the two are perpendicular to each other.
Of course, if you calculate the energy flow
$\mathbf{S}=\BE\times\BB=\frac{i}{2}\BF\times\bar{\BF}$
(where $\bar{\BF}$ denotes the complex conjugate of $\BF$),
you'll find that the direction of the energy flow doesn't
coincide with the $z$ direction of wave propagation,
which is of course true because the energy flow is always
perpendicular to the electromagnetic field.
Also, since this final solution generally has
$\BF^2=F_z^2+F_wF_{\bar{w}}$
with non-zero real and imaginary parts,
thus $\BE$ and $\BB$ are not orthogonal
in general,
and there is no $|\BE|=|\BB|$,
although both conclusions hold in the case of
vacuum plane waves.

Exactly similar to the above treatment,
from the equations (\ref{m1c1}), (\ref{m1c2}), (\ref{m1c3}), (\ref{m1c4})
we can also get some other solutions to Maxwell's equations,
for example, we can get the following solutions (which can be verified directly)
\begin{align}F_z&=B(u, w)+W(v, \bar{w}),\nonumber\\
 F_w&=B_w(u, w)v+B_u(u,w)\bar{w}+f(u, w),\nonumber\\
F_{\bar{w}}&=-W_{\bar{w}}(v,\bar{w})u-W_v(v, \bar{w})w+g(v,\bar{w}). \label{analytics2}\end{align}
where $B_w\equiv\p_wB, B_u\equiv\p_uB$, $W_{\bar{w}}\equiv\p_{\bar{w}}W, W_v\equiv\p_v W$.

 The solution (\ref{analytics1}) and the solution (\ref{analytics2}),
 and their linear superposition, exhausts all possibilities for
 $\p_u\p_vF_z=0,\ \p_w\p_{\bar{w}}F_z=0$.
 So what about the case of $\p_u\p_vF_z\neq 0,\ \p_w\p_{\bar{w}}F_z\neq 0$?
 This is what will be discussed in the next section.

\section{The spinor form of Maxwell's equations and several theorems on solving it}
\label{newform}

To discuss the more general case, we may wish to notate
$F_z=\Phi(v,u,w,\bar{w})$, and to summarize the
equations (\ref{m1c1}), (\ref{m1c2}), (\ref{m1c3}), (\ref{m1c4}) as follows,
\begin{align}&\p_vF_w=\p_w\Phi,\quad \p_{\bar{w}}F_w=\p_u\Phi
\label{geq1}\\
&\p_u F_{\bar{w}}=-\p_{\bar{w}}\Phi,\quad \p_wF_{\bar{w}}=-\p_v\Phi.\label{geq2}\end{align}
This is the form of the vacuum source-free
Maxwell's equations that we eventually arrive which is more easily solvable.

In fact, the equations (\ref{geq1}) and (\ref{geq2}) are the spinor
form of source-free Maxwell's equations derived by Penrose\cite{Penrose1},
just in a different notation.
To see this clearly, it is useful to rewrite Eq.(\ref{geq1}) and Eq.(\ref{geq2})
in the following matrix form \begin{align}\begin{pmatrix}\p_v & \p_w\\
\p_{\bar{w}} & \p_u\end{pmatrix}\begin{pmatrix}-F_w & \Phi\\
\Phi & F_{\bar{w}}\end{pmatrix}=0.\label{geqmatrix}\end{align}
Notice that \be\begin{pmatrix}\p_v & \p_w\\
\p_{\bar{w}} & \p_u\end{pmatrix}\sim \sigma^{\mu}\p_{\mu},\nonumber\ee
where the four components of the $\sigma^{\mu}=(1,\sigma^x,\sigma^y,\sigma^z)$
are $2\times 2$ unit matrix as well as three Pauli matrices, and $\p_{\mu}=(\p_t, \p_x, \p_y, \p_z)$.
Further define the symmetric double-spinor $\Psi$ as \be \Psi\equiv\begin{pmatrix}-F_w & \Phi\\
\Phi & F_{\bar{w}}\end{pmatrix}. \nonumber\ee
Then we can rewrite the equation (\ref{geqmatrix}) in standard spinor form as
\be\sigma^{\mu}\p_{\mu}\Psi=0.\ee

However, for the purposes of this article, it is still the equation (\ref{geq1}) and the equation (\ref{geq2}) that we really use. Of course, this set of equations is completely equivalent to Maxwell's equations, but it is so much easier to solve than solving Maxwell's equations directly. First of all, from both the two equations of (\ref{geq1}) and the two equations of (\ref{geq2}), it is easy to derive \be (\p_u\p_v-\p_w\p_{\bar{w}})\Phi=0\Leftrightarrow \Phi_{vu}=\Phi_{w\bar{w}}. \label{geq3}\ee Where $\Phi_{vu}=\Phi_{uv}\equiv\p_v\p_u\Phi$, $\Phi_{w\bar{w}}=\Phi_{\bar{w}w}\equiv\p_w\p_{\bar{w}}\Phi$. Since $4\p_v\p_u=\p^2_t-\p^2_z$ and $4\p_w\p_{\bar{w}}=\p^2_x+\p^2_y$, so of course the above equation is just the standard scalar D'Alembert equation $(\p^2_t-\p^2_x-\p^2_y-\p^2_z)\Phi=0$.

Since the case of $\p_u\p_v\Phi=\p_w\p_{\bar{w}}\Phi=0$ has been systematically discussed in the previous two sections,
we discuss below the case where $\p_u\p_v\Phi=\p_w\p_{\bar{w}}\Phi\neq 0$. We will prove the following \textbf{General Theorem:} Given any $\Phi(v,u,w,\bar{w})$ that satisfies the scalar D'Alembert equation $\p_u\p_v\Phi=\p_w\p_{\bar{w}}\Phi$,
we can obtain a set of solutions of (\ref{geq1}) and (\ref{geq2}) by simple indefinite integration as follows
\begin{align} & F_w=\int dv \Phi_w(v,u,w,\bar{w})+f(u, w)\label{gsolution1}\\
 & F_{\bar{w}}=-\int du \Phi_{\bar{w}}(v,u,w,\bar{w})+g(v, \bar {w}). \label{gsolution2}\end{align}
where $\Phi_w\equiv\p_w\Phi$, $\Phi_{\bar{w}}\equiv\p_{\bar{w}}\Phi$.

The proof is as follows: It may be useful to take (\ref{gsolution1}) as an example,
and we show that it satisfies both two equations of (\ref{geq1}),
and the proof that (\ref{gsolution2}) satisfies (\ref{geq2}) is exactly similar.
First of all it is obvious that (\ref{gsolution1}) satisfies $\p_vF_w=\p_w\Phi$.
In order to show that the (\ref{gsolution1}) also satisfies $\p_{\bar{w}}F_w=\p_u\Phi$, we note that
according to Eq.(\ref{gsolution1}), we have \be \p_{\bar{w}}F_w=\p_{\bar{w}}\int dv \Phi_w=
\int dv \Phi_{w\bar{w}}=\int dv \p_v\p_u\Phi=\p_u\Phi.\ee
where we have utilized the fact that $\Phi$ satisfies the scalar D'Alembert equation (\ref{geq3}).
This completes the proof. Of course, this proof only applies to the case where $\Phi_{vu}=\Phi_{w\bar{w}}\neq 0$.

To give an example, let's assume that the $v,u$ variables in $\Phi$ can be separated, i.e., setting \be \Phi(v,u,w,\bar{w})=e^{-ik_Lv}e^{-ik_Ru}\phi(w,\bar{w}). \ee
Where $k_L, k_R$ are two real numbers. Substituting into the scalar D'Alembert equation (\ref{geq3}), we have \be (\p_w\p_{\bar{w}}+k_Lk_R)\phi(w,\bar{w})=0.\ee
That is, $\phi(w,\bar{w})$ satisfies Helmholtz's equations
in the two-dimensional $(x,y)$ plane.

Then, according to the general theorem above,
we can get the following solutions of equations (\ref{geq1}) and (\ref{geq2}) \begin{align} &F_w=\frac{i}{k_L}e^{-ik_Lv}e^{-ik_Ru}\phi_w(w,\bar{w})+f(u, w)\label {hsolution1}\\
&F_{\bar{w}}=-\frac{i}{k_R}e^{-ik_Lv}e^{-ik_Ru}\phi_{\bar{w}}(w,\bar{w})+g(v, \bar{w}). \label{hsolution2}\end{align}
where $\phi_w\equiv\p_{w}\phi$, $\phi_{\bar{w}}\equiv\p_{\bar{w}}\phi$.
This example can certainly be used in solving the electromagnetic waves in a waveguide.
One might as well call the conclusion of this example the \textbf{Waveguide Solution Theorem}.

Even more wonderfully, by utilizing the general theorem, we can relate the solution of
Maxwell's equations to the solution of the two-dimensional Schrodinger equation. To do this, we take
\be \Phi(v,u,w,\bar{w})=e^{-i\frac{k}{2}v}\psi(u, w,\bar{w}). \ee
where $k$ is a real number. Substituting into the scalar D'Alembert equation (\ref{geq3}), we have
\be i\p_u\psi(u,w,\bar{w})=-\frac{2}{k}\p_{w}\p_{\bar{w}}\psi(u,w,\bar{w})=-\frac{1}{2k}(\p^2_x+\p^2_y)\psi.
\label{schrondinger1}\ee
This is the Schrodinger equation (taking $\hbar=1$) for a two-dimensional free particle with mass $k$.
It certainly has many analytical solutions.

Using the general theorem, we can obtain the following \textbf{Schrodinger Solution Theorem}:
Given any solution $\psi(u,w,\bar{w})$ to the equation (\ref{schrondinger1}),
the following set of solutions to the equations (\ref{geq1}) and (\ref{geq2}) can be obtained
\begin{align} &F_w=2\frac{i}{k}e^{-i\frac{k}{2}v}\psi_w(u, w,\bar{w})+f(u, w)\label{schrondinger2}\\
&F_{\bar{w}}=-e^{-i\frac{k}{2}v}\int du \psi_{\bar{w}}(u, w,\bar{w})+g(v, \bar{w}). \label{schrondinger3}\end{align}
where $\psi_w\equiv\p_w\psi$, $\psi_{\bar{w}}\equiv\p_{\bar{w}}\psi$.

Let us conclude this section by reviewing again the equation (\ref{geq1}), and the equation (\ref{geq2}),
and we find that together they look a little like the Cauchy-Riemann equations for two-dimensional complex analysis.
Perhaps (\ref{geq1}), (\ref{geq2}) can be seen as a generalization of the Cauchy-Riemann equations in four-dimensional space-time. If that is the case, then their solutions should have some wonderful and profound mathematical properties,
just as the solutions of the Cauchy-Riemann equations have wonderful and profound mathematical properties.
In this way, the solutions with the wonderful mathematical properties that we found in the previous two sections are not difficult to understand.

\section{Spacetime diffraction formula}
\label{SDF}

In this section we specialize in applications of Schrodinger solution theorem.

For example, take the following analytic solution to the equation (\ref{schrondinger1}) \be\psi(u,w,\bar{w})=\frac{A}{u}\exp\big(\frac{ik}{2u}|w|^2\big). \label{prop0}\ee
where $A$ is an arbitrary complex constant. Then, according to the equation (\ref{schrondinger2})
we can get \be F_w=-\frac{A}{u^2}e^{-i\frac{k}{2}v}\bar{w}\exp\big(\frac{ik}{2u}|w|^2\big)+f(u, w). \ee
And according to the equation (\ref{schrondinger3}) we can get
\begin{align}F_{\bar{w}}&=-A e^{-i\frac{k}{2}v}\int du \frac{ik}{2u^2}w\exp\big(\frac{ik}{2u}|w|^2\big)+g (v, \bar{w})\nonumber\\
&=\frac{A}{\bar{w}}e^{-i\frac{k}{2}v}\exp\big(\frac{ik}{2u}|w|^2\big)+g(v, \bar{w}). \end{align}
Of course, if you're willing to write the solution of $F_z$ at the same time,
it's \be F_z=\Phi(v,u,w,\bar{w})=\frac{A}{u}e^{-i\frac{k}{2}v}\exp\big(\frac{ik}{2u}|w|^2\big). \ee
This solution to Maxwell's equations might be difficult to obtain by other methods, but,
using Schrodinger solution theorem,
we easily obtain it. Of course, this solution fails at $u=0$.

As for taking other analytic solutions of Schrodinger's equation
(\ref{schrondinger1}), using Schrodinger solution theorem,
we can also obtain a wide variety of exotic solutions of Maxwell's equations,
which we will not discuss in this paper!

The solution discussed above
has an important application.
Indeed, if one knows the initial value of the Schrodinger equation (\ref{schrondinger1})
for the wave function at $u=0$, which may be written as $\psi_0(w,\bar{w})\equiv\psi(0,w,\bar{w})$,
then we can use the quantum mechanical propagator method
to find the wave function $\psi(u,w,\bar{w})$ at any $u$.
Remembering that the corresponding quantum mechanical propagator is $K(u,w,\bar{w})$,
we have \be \psi(u,w,\bar{w})=\int d^2w'K(u-0,w-w',\bar{w}-\bar{w}')\psi_0(w',\bar{w}'),\ee
where $d^2w'=dx'dy'$, and the integral spreads over the entire $(x',y')$ plane.
According to standard quantum mechanical knowledge \cite{Feynman2},
the propagator $K(u,w,\bar{w})$ here is \be K(u,w,\bar{w})=\frac{k}{2\pi i u}\exp\big(\frac{ik}{2u}|w|^2\big). \ee
Further, we can follow the treatment of Schrodinger solution theorem to obtain the
\textbf{Spacetime Diffraction Formula} for Maxwell's equations, which, of course,
is now an exact solution that
is an exact reconstruction of the electromagnetic wave throughout space-time
from the initial value of $F_z$ in the diffraction plane at $u=0$.

Specifically, we still have $F_z=\Phi(v,u,w,\bar{w})=e^{-i\frac{k}{2}v}\psi(u, w,\bar{w})$,
where \be\psi(u,w,\bar{w})=\int d^2w'\frac{k}{2\pi i u}\exp\big( \frac{ik}{2u}|w-w'|^2\big)\psi_0(w',\bar{w}'). \ee
Substituting into equations (\ref{schrondinger2}) and (\ref{schrondinger3}), we can get
\begin{align} F_w&=\frac{ik}{2\pi u^2}e^{-i\frac{k}{2}v}\int d^2w'(\bar{w}-\bar{w}')\exp\big(\frac{ik}{2u}|w-w'|^2\big)\psi_0(w',\bar{w}'). \nonumber\\
F_{\bar{w}}&=\frac{k}{2\pi i}e^{-i\frac{k}{2}v}\int d^2w'\frac{1}{\bar{w}-\bar{w}'}\exp\big(\frac{ik}{2u}|w-w'|^2\big)\psi_0(w',\bar {w}'). \end{align} Where we have taken the undefined $f(u, w)=0, g(v, \bar {w})=0$.
The reason why we can take both of these to be zero is as follows (taking the example of why $f(u, w)=0$):
firstly, since there is no source on the whole observation screen, $f(u, w)$ must be holomorphic on the whole complex plane
of the observation screen, which can only be the case if $f(u, w)=f(u)$, i.e., it depends only on $u$, but $f(u)$ describes a
right moving background plane wave, and this background plane wave does not exist in practice,
so it is only possible that $f(u, w)=f(u)=0$. Similarly, it can be argued that $g(v,\bar{w})=0$.

The above result is similar to the textbook Fresnel diffraction formula \cite{optics},
but now the result is an exact result and it is used to reconstruct
the electromagnetic field in the whole spacetime,
not just in space, so we call it \textbf{Spacetime Diffraction Formula}.

\section{Contour Integrals and Two Simple Theorems on Electromagnetic Fields}

In the the previous section, we found that the source-free Maxwell's equations can be rewritten in the following form \begin{align}&\p_vF_w=\p_w\Phi,\quad \p_u F_{\bar{w}}=-\p_{\bar{w}}\Phi \label{cgeq1}\\\
&\p_{\bar{w}}F_w=\p_u\Phi,\quad \p_wF_{\bar{w}}=-\p_v\Phi.\label{cgeq2}\end{align}
As already mentioned in section (\ref{newform}), 
this form is somewhat like a generalization of the Cauchy-Riemann equations in complex analysis to four-dimensional space-time. This motivates us to apply some concepts from complex analysis here, but of course, now only the $(x,y)$ plane (that is, the complex $w$ plane) is the complex plane, and the concepts of complex analysis can only be used in this plane.

For example, by taking any closed contour $L$ in a simply connected source-free region $D$ in the complex $w$-plane, we can examine the contour integral $\oint_L F_wdw$ as well as $\oint_L F_{\bar{w}}d\bar{w}$, and, of course, $F_w$ does not necessarily have to be a holomorphic function on $D$, and $F_{\bar{w}}$ does not necessarily have to be an anti holomorphic function, so neither of these two contour integrals is necessarily equal to zero. But we have the following theorem:

\textbf{Theorem 1:} The contour integrals $\oint_L F_wdw$ and $\oint_LF_{\bar{w}}d\bar{w}$ on a simply connected source-free region $D$ of the complex $w$-plane satisfy the following relation
\be \p_v\oint_L F_wdw=\p_u\oint_L F_{\bar{w}}d\bar{w}. \label{result1}\ee

Proof: Examining $\p_v\oint_L F_wdw-\p_u\oint_L F_{\bar{w}}d\bar{w}$, using the fundamental equation (\ref{cgeq1}), we have \begin{align}&\p_v\oint_L F_wdw-\p_u\oint_L F_{\bar{w} }d\bar{w}=
\oint_L \p_vF_wdw-\oint_L \p_uF_{\bar{w}}d\bar{w}\nonumber\\
=&\oint_L \big(\p_w\Phi dw+\p_{\bar{w}}\Phi d\bar{w}\big)=0.\end{align}
The final equality to zero is due to the fact that $\p_w\Phi dw+\p_{\bar{w}}\Phi d\bar{w}$
is a total derivative in the $(x, y)$ plane. The proof is finished.

We give just one specific example to further test the above theorem.
In the previous section (\ref{SDF}), we gave the following solution to the source-free Maxwell's equations,
\begin{align} F_w&=-\frac{A}{u^2}e^{-i\frac{k}{2}v}\bar{w}\exp\big(\frac{ik}{2u}|w|^2\big)\nonumber\\
F_{\bar{w}}&=\frac{A}{\bar{w}}e^{-i\frac{k}{2}v}\exp\big(\frac{ik}{2u}|w|^2\big)\nonumber\\
\Phi &=\frac{A}{u}e^{-i\frac{k}{2}v}\exp\big(\frac{ik}{2u}|w|^2\big). \end{align}
By using this solution, and using the standard Stokes formula (noting that $S$ is the area enclosed
by the contour $L$ in the $w$-plane), it is not difficult to get
\begin{align}&\oint_L F_wdw=-\int_S\p_{\bar{w}}F_w dw\wedge d\bar{w}\nonumber\\
=&\frac{A}{u^2}e^{-i\frac{k}{2}v}
\int_S\big(1+\frac{ik}{2u}|w|^2\big)\exp\big(\frac{ik}{2u}|w|^2\big)dw\wedge d\bar{w}. \label{verify0}\end{align}
Similarly
\begin{align}&\oint_L F_{\bar{w}}d\bar{w}=\int_S\p_wF_{\bar{w}}dw\wedge d\bar{w}\nonumber\\
=& A e^{-i\frac{k}{2}v}\frac{ik}{2u}\int_S\exp\big(\frac{ik}{2u}|w|^2\big)dw\wedge d\bar{w}.\end{align}
Now it is not difficult to directly verify
\begin{align}\p_v\oint_L F_wdw &=-A\frac{ik}{2u^2}e^{-i\frac{k}{2}v}
\int_S\big(1+\frac{ik}{2u}|w|^2\big)\exp\big(\frac{ik}{2u}|w|^2\big)dw\wedge d\bar{w}\nonumber\\
&=\p_u\oint_L F_{\bar{w}}d\bar{w}. \end{align}

The above example motivates us to directly examine $\oint_L F_wdw$ and $\oint_L F_{\bar{w}}d\bar{w}$ itself, which in turn leads to the following theorem:

\textbf{Theorem 2:} The contour integrals $\oint_L F_wdw$ and $\oint_L F_{\bar{w}}d\bar{w}$ on a simply connected source-free region $D$ of the complex $w$-plane satisfy the following relations, respectively
\begin{align}\frac{1}{2\pi i}\oint_L F_wdw &=\frac{1}{\pi}\p_u\int_S\Phi d^2w.\label{theorem21}\\
\frac{1}{2\pi i}\oint_L F_{\bar{w}}d\bar{w}&=\frac{1}{\pi}\p_v\int_S\Phi d^2w.\label{theorem22}\end{align}
where $d^2w\equiv \frac{i}{2}dw\wedge d\bar{w}=dx\wedge dy$.

Proof: Taking the proof of (\ref{theorem21}) as an example, the proof of (\ref{theorem22}) is exactly similar. Using the fundamental equation (\ref{cgeq2}), we have
\begin{align}&\frac{1}{2\pi i}\oint_L F_wdw =
-\frac{1}{2\pi i}\int_S \p_{\bar{w}}F_wdw\wedge d\bar{w}\nonumber\\\
=&\frac{i}{2\pi}\int_S\p_u\Phi dw\wedge d\bar{w}=\frac{1}{\pi}\p_u\int_S\Phi d^2w.\end{align}
Theorem proved.

Obviously, Theorem 1 can serve as a corollary to Theorem 2. But since logically it can be proved independently, we list it separately.

We still use the example used above to further test this theorem,
taking the verification of formula (\ref{theorem21}) as an example,
for which it is sufficient to carry out the following calculations (substituting the solution of $\Phi$ given in the specific example above)
\begin{align}&\p_u\int_S\Phi d^2w=\int_S\p_u\Phi d^2w\nonumber\\\
=&-\frac{A}{u^2} e^{-i\frac{k}{2}v}\int_S d^2w\big(1+\frac{ik}{2u}|w|^2\big)\exp\big(\frac{ik}{2u}|w|^2\big). \end{align}
Comparing this result with (\ref{verify0}) above, it is easy to see that the theorem holds.

The proofs of both theorems are very simple, but the theorems themselves are very nice and hold for any electromagnetic wave in source-free region, so they have their independent applications! And might make some sense mathematically, too.

\section{Acknowledgements}
The authors acknowledge the support of the Science and Technology Project of Department of Education of Jiangxi Province under Grant No. GJJ180378

\end{document}